\documentclass[english]{article}
\usepackage[T1]{fontenc}
\usepackage[latin1]{inputenc}
\usepackage{geometry}
\usepackage{amsmath}
\usepackage{setspace}
\usepackage{amssymb}

\makeatletter
\newcommand{\lyxaddress}[1]{
\par {\raggedright #1
\vspace{1.4em}
\noindent\par}
}

\usepackage{babel}
\makeatother
\begin{document}

\title{Quaternionic Reformulation of Generalized Superluminal Electromagnetic
Fields of Dyons}

\author{P. S. Bisht, Jivan Singh and O. P. S. Negi%
\thanks{Address from July 01 to September 15, 2007:- Institute of Theoretical
Physics, Chinese Academy of Sciences , KITP Building Room No.- 6304,
Hai Dian Qu Zhong Guan Chun Dong Lu, 55 Hao , Beijing - 100080, P.R.China%
}}

\maketitle

\lyxaddress{\begin{center}
Department of Physics\\
Kumaun University\\
S. S. J. Campus\\
Almora - 263601 (India)
\par\end{center}}

\lyxaddress{\begin{center}
Email:- ps\_bisht123@rediffmail.com\\
jgaria@indiatimes.com\\
ops\_negi@yahoo.co.in
\par\end{center}}

\begin{abstract}
Superluminal electromagnetic fields of dyons are described in $T^{4}$-
space and Quaternion formulation of various quantum equations is derived.
It is shown that on passing from subluminal to superluminal realm
via quaternion the theory of dyons becomes the Tachyonic dyons. Corresponding
field Equations of Tachyonic dyons are derived in consistent, compact
and simpler form.


\end{abstract}

\section{Introduction}

The question of existence of monopoles (dyons) \cite{key-1,key-2,key-3}
has become a challenging new frontier and the object of more interest
in connection with the current grand unified theories \cite{key-4,key-5},
quark confinement problem of quantum chromo dynamics \cite{key-6},
the possible magnetic condensation of vacuum \cite{key-7}, their
role as catalyst in proton decay \cite{key-8} and supersymmetry \cite{key-9,key-10}.
The eight decades of this century witnessed a rapid development of
the group theory and gauge field theory to establish the theoretical
existence of monopoles and to explain their group properties and symmetries.
Keeping in mind t' Hooft-Polyakov and Julia-Zee solutions \cite{key-11,key-12}
and the fact that despite the potential importance of monopoles, the
formalism necessary to describe them has been clumsy and not manifestly
covariant, a self consistent quantum field theory of generalized electromagnetic
fields associated with dyons (particle carrying electric and magnetic
charges) of various spins has been developed \cite{key-13,key-14}
by introducing two four potentials \cite{key-15} and avoiding string
variables \cite{key-16} with the assumption of the generalized charge,
generalized four-potential, generalized field tensor, generalized
vector field and generalized four-current density associated with
dyons as complex quantities with their real and imaginary parts as
electric and magnetic constituents. On the other hand, there has been
continuing interest \cite{key-17,key-18,key-19,key-20} in higher
dimensional kinematical models for proper and unified theory of subluminal
(bradyon) and superluminal (tachyon) objects \cite{key-21,key-22}.
The problem of representation and localizations of superluminal particles
has been solved only by the use of higher dimensional space \cite{key-23,key-24,key-25}
and it has been claimed that the localization space for tachyons is
$T^{4}$- space with one space and three times while that for bradyon
is $R^{4}$- space in view of localizability and of these particles. 

Introducing the concepts of superluminal Lorentz transformations,
need of higher dimensional space-time, localizability of bradyons
and tachyons, in the present paper, we have under taken the study
of generalized fields of dyons under superluminal Lorentz transformations
($SLTs$). It has been shown that the generalized electromagnetic
fields behave in frame $K'$ (i.e. the superluminal frame) as subluminal
fields do in frame $K$ (subluminal frame). As such, the generalized
fields, when viewed upon by an observer in bradyonic frame, appear
as superluminal fields and thus, satisfy the field equations different
from Maxwell's equations used for electric charge ( or magnetic monopole)
and generalized Dirac-Maxwell's (GDM) equations of dyons. Hence, it
is concluded that the superluminal electromagnetic fields are not
same as the familiar electric and magnetic fields associated with
electric charge (or magnetic monopole) and dyons of our every day
experience. It is shown that the superluminal electromagnetic fields
and field equations are no more invariant under SLTs with the chronological
mapping of space-time on passing from subluminal to superluminal realm.
It is has been emphasized that in order to retain the Lorentz invariance
of field equations, we are forced to include extra negative sign to
the components of four-current densities for electric charge (or magnetic
monopole) and dyons respectively. It is also concluded that though
the roles of electric and magnetic charges are not changed while passing
from subluminal to superluminal realm under $SLTs$, a dyon interacting
with superluminal electromagnetic fields behaves neither as electric
charge, nor as pure magnetic monopole but having the mixed behaviour
of electric and magnetic charges, rather, namely a tachyonic dyon.
Describing the need of higher dimensional and localizability spaces
for bradyons and tachyons respectively as $R^{4}-$and $T^{4-}$ spaces,
we have obtained superluminal electromagnetic fields in $T^{4}-$spaces
and derived the consistent and manifestly covariant field equations
and equation of motion where the velocity is described as reversed
velocity. Starting with the quaternionic form of generalized four-potential
of dyons, we have developed the simple and compact quaternionic form
of Maxwell's equations and it has been shown that while passing from
usual four space to quaternionic formulation the signature of four-vector
is changed from $(+,-,-,-)$ to $(-,-,-,+)$. Hence, the quaternionic
formulation and superluminal behaviour have striking similarities.
The corresponding quaternionic field equations of bradyonic and tachyonic
dyons are derived consistently in $R^{4}-$ and $T^{4}-$ spaces respectively
in consistent, simple and compact formulations. These quaternionic
formulations reproduce the theories of electric (magnetic) charge
in the absence of magnetic (electric) charge or vice versa on dyons
in $R^{4}-$and $T^{4}-$ spaces.

\section{Field Associated with Dyons}

Let us define the generalized charge on dyons as \cite{key-13,key-14}

\begin{eqnarray}
q & = & e-ig\,\,(i=\sqrt{-1})\label{eq:1}\end{eqnarray}
where $e$ and $g$ are respectively electric and magnetic charges.
Generalized four - potential $\{ V_{\mu}\}=\{\phi,\overrightarrow{V}\}$associated
with dyons is defined as,

\begin{eqnarray}
V_{\mu} & = & A_{\mu}-iB_{\mu}\label{eq:2}\end{eqnarray}
where $\{ A_{\mu}\}=\{\phi_{e},\overrightarrow{A}\}$ and $\{ B_{\mu}\}=\{\phi_{g},\overrightarrow{B}\}$
are respectively electric and magnetic four - potentials. We have
used throughout the natural units $c=\hbar=1$ and Minikowski space
id described with the signature $(-,+,+,+)$. Generalized electric
and magnetic fields of dyons are defined in terms of components of
electric and magnetic potentials as,

\begin{eqnarray}
\overrightarrow{E} & = & -\frac{\partial\overrightarrow{A}}{\partial t}-\overrightarrow{\nabla}\phi_{e}-\overrightarrow{\nabla}\times\overrightarrow{A},\nonumber \\
\overrightarrow{H} & = & -\frac{\partial\overrightarrow{B}}{\partial t}-\overrightarrow{\nabla}\phi_{g}+\overrightarrow{\nabla}\times\overrightarrow{B}.\label{eq:3}\end{eqnarray}
These electric and magnetic fields of dyons are invariant under generalized
duality transformation i.e. 

\begin{eqnarray}
A_{\mu} & = & A_{\mu}\cos\theta+B_{\mu}\sin\theta\nonumber \\
B_{\mu} & = & A_{\mu}\sin\theta+B_{\mu}\cos\theta.\label{eq:4}\end{eqnarray}
The expression of generalized electric and magnetic field given by
equation (\ref{eq:3}) are symmetrical and both the electric and magnetic
field of dyons may be written in terms of longitudinal and transverse
components. The generalized field vector $\overrightarrow{\psi}$
associated with dyons is defined as

\begin{eqnarray}
\overrightarrow{\psi} & = & \overrightarrow{E}-i\overrightarrow{H}\label{eq:5}\end{eqnarray}
and accordingly, we get the following differential form of generalized
Maxwell's equations for dyons i.e.

\begin{eqnarray}
\overrightarrow{\nabla}.\overrightarrow{\psi} & = & J_{0},\nonumber \\
\overrightarrow{\nabla}\times\overrightarrow{\psi} & = & -i\overrightarrow{J}-i\frac{\partial\overrightarrow{\psi}}{\partial t}\label{eq:6}\end{eqnarray}
where $J_{0}$ and $\overrightarrow{J}$, are the generalized charge
and current source densities of dyons, given by;

\begin{eqnarray}
J_{\mu} & = & j_{\mu}-ik_{\mu}\equiv\{ J_{0},\overrightarrow{J}\}.\label{eq:7}\end{eqnarray}
Substituting relation (\ref{eq:3}) into equation (\ref{eq:5}) and
using equation (\ref{eq:2}), we obtain the following relation between
generalized field vector and generalized potential of dyons i.e.

\begin{eqnarray}
\overrightarrow{\psi} & = & -\overrightarrow{\nabla}\phi-\frac{\partial\overrightarrow{V}}{\partial t}-i\overrightarrow{\nabla}\times\overrightarrow{V}.\label{eq:8}\end{eqnarray}
In equation (\ref{eq:8}) , $\{ j_{\mu}\}=(\rho_{e},\overrightarrow{j})$
and $\{ k_{\mu}\}=(\rho_{g},\overrightarrow{k})$ are electric and
magnetic four current densities. Thus we write the following tensor
forms of generalized Maxwell's - Dirac equations of dyons \cite{key-13,key-14}
i.e.

\begin{eqnarray}
F_{\mu\nu,\nu} & = & j_{\mu}\nonumber \\
\widetilde{F_{\mu\nu,\nu}} & = & k_{\mu}\label{eq:9}\end{eqnarray}

where

\begin{eqnarray}
F_{\mu\nu} & = & E_{\mu\nu}-\widetilde{H_{\mu\nu}},\nonumber \\
\widetilde{F_{\mu\nu}} & = & H_{\mu\nu}+\widetilde{E_{\mu\nu}}\label{eq:10}\end{eqnarray}

with

\begin{eqnarray}
E_{\mu\nu} & = & \partial_{\nu}A_{\mu}-\partial_{\mu}A_{\nu},\nonumber \\
H_{\mu\nu} & = & \partial_{\nu}B_{\mu}-\partial_{\mu}B_{\nu},\nonumber \\
\widetilde{E_{\mu\nu}} & = & \frac{1}{2}\varepsilon_{\mu\nu\rho\lambda}E^{\rho\lambda},\nonumber \\
\widetilde{H_{\mu\nu}} & = & \frac{1}{2}\varepsilon_{\mu\nu\rho\lambda}H^{\rho\lambda}.\label{eq:11}\end{eqnarray}
The tidle $(\thicksim)$ denotes the dual part while $\varepsilon_{\mu\nu\rho\lambda}$
are four index Levi - Civita symbol. Generalized fields of dyons given
by equations (\ref{eq:3}) may directly be obtained from field tensors
$F_{\mu\nu}$ and $F_{\mu\nu}^{d}$ as,

\begin{eqnarray}
F_{0i} & = & E^{i},\nonumber \\
F_{ij} & = & \varepsilon_{ijk}H^{k},\nonumber \\
H_{0i}^{d} & = & -H^{i},\nonumber \\
H_{ij}^{d} & = & -\varepsilon_{ijk}E^{k}.\label{eq:12}\end{eqnarray}

Taking the curl of second equation of (\ref{eq:6}) and using first
equation of (\ref{eq:6}), we obtain a new vector parameter $\overrightarrow{S}$
(say) i.e.

\begin{eqnarray}
\overrightarrow{S} & =\square\overrightarrow{\psi} & =-\frac{\partial\overrightarrow{\psi}}{\partial t}-i\overrightarrow{\nabla}\times\overrightarrow{J}-\overrightarrow{\nabla}\rho\label{eq:13}\end{eqnarray}
where $\square$ represents the D'Alembertian operator i.e.

\begin{eqnarray}
\square & =\nabla^{2}-\frac{\partial^{2}}{\partial t^{2}} & =\frac{\partial^{2}}{\partial x^{2}}+\frac{\partial^{2}}{\partial y^{2}}+\frac{\partial^{2}}{\partial z^{2}}-\frac{\partial^{2}}{\partial t^{2}}.\label{eq:14}\end{eqnarray}
Defining generalized field tensor as

\begin{eqnarray}
G_{\mu\nu} & = & F_{\mu\nu}-i\,\widetilde{F_{\mu\nu}}\label{eq:15}\end{eqnarray}
one can directly obtain the following generalized field equation of
dyons i.e.

\begin{eqnarray}
G_{\mu\nu,\nu} & = & J_{\mu},\nonumber \\
\widetilde{G_{\mu\nu,\nu}} & = & 0\label{eq:16}\end{eqnarray}
where $G_{\mu\nu}=V_{\mu,\nu}-V_{\nu,\mu}$ is called the generalized
electromagnetic field tensor of dyons. Equation (\ref{eq:16}) may
also be written as follows like second order Klein-Gordon equation
for dyonic fields

\begin{eqnarray}
\square V_{\mu} & = & J_{\mu}\,\qquad(in\, Lorentz\, gauge)\label{eq:17}\end{eqnarray}
Equations ( \ref{eq:9}) and (\ref{eq:16}) are also invariant under
duality transformations;

\begin{eqnarray}
(F,F^{d}) & = & (F\cos\theta+F^{d}\sin\theta,F\sin\theta-F^{d}\cos\theta),\label{eq:18}\\
(j_{\mu},k_{\mu}) & = & (j_{\mu}\cos\theta+k_{\mu}\sin\theta,j_{\mu}\sin\theta-k_{\mu}\cos\theta)\label{eq:19}\end{eqnarray}
where

\begin{eqnarray}
\frac{g}{e} & = & \frac{B_{\mu}}{A_{\mu}}=\frac{k_{\mu}}{j_{\mu}}=-\tan\theta=constant.\label{eq:20}\end{eqnarray}
Consequently the generalized charge of dyons may be written as 

\begin{eqnarray}
q & = & |q|e^{-i\theta}.\label{eq:21}\end{eqnarray}
The suitable Lagrangian density, which yields the field equation (\ref{eq:16})
under the variation of field parameters i.e. potential only without
changing the trajectory of particle, may be written as follows;

\begin{eqnarray}
\mathcal{L} & = & -m_{0}-\frac{1}{4}G_{\mu\nu}G_{\mu\nu}^{\star}+V_{\mu}^{\star}J_{\mu}\label{eq:22}\end{eqnarray}
where $m_{0}$ is the rest mass of the particle and $(\star)$ denotes
the complex conjugate. Lagrangian density given by equation (\ref{eq:22})
directly follows the following form of Lorentz force equation of motion
for dyons i.e.

\begin{eqnarray}
f_{\mu} & = & m_{0}\ddot{x}_{\mu}=Re\,(q^{\star}G_{\mu\nu})u^{\nu}\label{eq:23}\end{eqnarray}
where $Re$ denotes real part, $\{\ddot{x}_{\mu}\}$ is the four -
acceleration and $\{ u^{\nu}\}$ is the four - velocity of the particle.

\section{Dyonic field Equations under Superluminal Lorentz Transformation}

Special theory of relativity has been extended in a straight forward
manner to superluminal inertial frames and it has been shown that
the existence of tachyons (particles moves faster than light) does
not violate the theory of relativity while their detection may require
a modification in certain established motion of causality. In deriving
Superluminal Lorentz transformation with relative velocity between
two frames greater than velocity of light, two main approaches are
adopted by different authors. In the first one adopted by Recami et
al \cite{key-26}, the components of a four vector field in the direction
perpendicular to relative motion become imaginary on passing from
subluminal to superluminal realm while in the second approach followed
by Antippa-Everett \cite{key-27} the real superluminal Lorentz transformation
are used. In the light of Gorini's theorem \cite{key-28} and the
conclusion and Pahor and Strnad \cite{key-29}, that with the real
transformations either the speed of light is not invariant or relative
velocity between the frames does not have a meaning, the superluminal
Lorentz transformation of Racami et al \cite{key-26} are closer to
the spirit of relativity in comparison to the real ones . In order
to examine the invariance of generalized Maxwell's equations of dyons
under imaginary superluminal Lorentz transformations \cite{key-26},
let us introduce two inertial frames $K$ and $K'$ whose axes are
parallel and whose origins coincide at $t=t\,'=0$. Let $K'$ moves
with respect to $K$ with a superluminal velocity $v>1$ directed
along $Z$ - axis, the transformation equations between the coordinates
of an event as seen in $K'$ and those of the same event in $K$,
may be written as follows \cite{key-26},

\begin{eqnarray}
x'_{j} & = & \pm ix_{j},\,\,(j=1,2)\nonumber \\
x'_{3} & = & \pm\gamma(x_{3}-vt),\,\,(v>1)\nonumber \\
t' & = & \pm\gamma(t-xv_{3})\label{eq:24}\end{eqnarray}
where

\begin{eqnarray}
\gamma & = & (v^{2}-1)^{-\frac{1}{2}}.\label{eq:25}\end{eqnarray}
From these transformations we have 

\begin{eqnarray}
-(t')^{2}+(x')_{1}^{2}+(x')_{2}^{2}+(x')_{3}^{3} & = & t^{2}-x_{1}^{2}-x_{2}^{2}-x_{3}^{2}\label{eq:26}\end{eqnarray}
which shows that the reference metric $(-1,-+1,+1,+1)$ in frame $K$
is transformed to the metric $(1,-1,-1,-1)$in frame $K'$ with the
transformations (\ref{eq:24}) and the roles of space and time get
interchanged for superluminal transformations. In other words, the
superluminal transformations lead to chronological mapping \cite{key-23,key-24,key-26,key-30,key-17},

\begin{eqnarray}
(3,1) & \longleftrightarrow & (1,3)\label{eq:27}\end{eqnarray}
for the components of space and time on passing from subluminal to
superluminal realm or vice versa and thus describes

\begin{eqnarray}
(x,y,z,t) & \rightarrow & (t',ix',iy',iz')\label{eq:28}\end{eqnarray}
from which we get

\begin{eqnarray}
\square & = & -\square'\label{eq:29}\end{eqnarray}
and the mapping\begin{eqnarray}
(\overrightarrow{\nabla},i\frac{\partial}{\partial t}) & \rightarrow & (\frac{\partial}{\partial t'},i\overrightarrow{\nabla}).\label{eq:30}\end{eqnarray}
Similar superluminal transformations may be derived for the components
of four potentials (electric, monopole, dyon) and we may obtain that

\begin{eqnarray}
|A_{\mu}'|^{2} & = & -|A_{\mu}|^{2},\nonumber \\
|B_{\mu}'|^{2} & = & -|B_{\mu}|^{2},\nonumber \\
|V_{\mu}'|^{2} & = & -|V_{\mu}|^{2}\label{eq:31}\end{eqnarray}
with the correspondence $(3,1)\longleftrightarrow(1,3)$ mapping we
get

\begin{eqnarray}
(A_{1},A_{2},A_{3},i\phi_{e}) & \rightarrow & (\phi_{e}',A_{1}',A_{2}',A_{3}'),\nonumber \\
(B_{1},B_{2},B_{3},i\phi_{g}) & \rightarrow & (\phi_{g}',B_{1}',B_{2}',B_{3}'),\nonumber \\
(V_{1},V_{2},V_{3},i\phi) & \rightarrow & (\phi',V_{1}',V_{2}',V_{3}').\label{eq:32}\end{eqnarray}
Using relation (\ref{eq:26}-\ref{eq:32}) and the similar mapping
for the component four current densities (electric, monopole, dyon)
we may transform the Maxwell's equation given by equation (\ref{eq:17})
in frame $K$ to the following equation in frame $K'$ i.e.;

\begin{eqnarray}
\square'A_{\mu}' & = & -j_{\mu}',\nonumber \\
\square'B_{\mu}' & = & -k_{\mu}',\nonumber \\
\square'V_{\mu}' & = & -J_{\mu}',\label{eq:33}\end{eqnarray}
which are the equations according to which the superluminal electromagnetic
fields associated respectively with electric charge , magnetic monopole
and dyon are coupled to their tachyonic counterparts which may be
considered as bradyons in superluminal frame $K'$ in view of tachyon-bradyon
reciprocity and, therefore, these electromagnetic fields behave in
frame $K'$ as the subluminal fields do in frame $K$. These fields,
when viewed upon by an observer in frame $K$ (Bradyonic frame), appear
as superluminal electromagnetic fields and satisfy the field equations
(\ref{eq:33}), which differs from the usual filed equations respectively
for electric charge, magnetic monopole and dyon. As such, it may be
concluded that the superluminal electromagnetic fields are not the
same as the familiar electric and magnetic fields of electric charge,
magnetic monopole and dyons of our daily experience and obey Maxwell
type equations in subluminal frame of reference. Consequently the
field equations are no more invariant under imaginary superluminal
transformations and to retain the Lorentz invariance of field equations
we are forced to include extra negative sign to the components of
four current densities for electric charge, magnetic monopole and
dyon respectively with incorporating the following mappings;

\begin{eqnarray}
(j_{1},j_{2},j_{3},i\rho_{e}) & \rightarrow & -(\rho_{e}',j_{1}',j_{2}',j_{3}'),\nonumber \\
(k_{1},k_{2},k_{3},i\rho_{g}) & \rightarrow & -(\rho_{g}',k_{1}',k_{2}',k_{3}'),\nonumber \\
(J_{1},J_{2},J_{3},i\rho) & \rightarrow & -(\rho',J_{1}',J_{2}',J_{3}').\label{eq:34}\end{eqnarray}
Despite the change in sign, the real and imaginary components of four
current densities lead to corresponding real and imaginary components
of the four potentials. The change in sign of charge and current densities
leaves the total charge invariant as the volume element also changes
the sign under imaginary superluminal Lorentz transformations. This
change of sign in the components of four current densities may lead
to the conclusion that the filed equations may be treated as invariant
on passing from subluminal to superluminal realm or vice versa \cite{key-31}.
If we use the mappings given by equations (\ref{eq:27},\ref{eq:28},\ref{eq:30}
and \ref{eq:32}) the generalized electric and magnetic fields of
dyons for superluminal case take the following expressions for generalized
superluminal electromagnetic fields;

\begin{eqnarray}
\overrightarrow{E}' & = & -grad'\phi_{e}'-\frac{\partial\overrightarrow{A}'}{\partial t'}-\frac{\partial}{\partial t'}\phi_{g}'\hat{n_{g},}\nonumber \\
\overrightarrow{H}' & = & -grad'\phi_{g}'-\frac{\partial\overrightarrow{B}'}{\partial t'}-\frac{\partial}{\partial t'}\phi_{e}'\hat{n_{e}}\label{eq:35}\end{eqnarray}
where $\hat{n}_{e}$ and $\hat{n}_{g}$ are unit vectors in the direction
of electric and magnetic fields associated with electric and magnetic
charges. These equations are different from those obtained earlier
by Negi-Rajput \cite{key-23} derived for electric charge only. These
are also not exactly same as given by equations (\ref{eq:3}) for
generalized subluminal electric and magnetic fields of dyons but shows
the striking symmetry between the electric and magnetic fields of
dyons under superluminal transformations and may thus be visualized
as the generalized superluminal electromagnetic field of dyons in
frame $K'$ when viewed from subluminal frame $K$. As such, it may
be concluded that though the roles of electric and magnetic charges
are not changed while passing from subluminal to superluminal realm
under the superluminal transformations, a dyon interacting with superluminal
electromagnetic fields containing symmetrical electromagnetic fields
behaves neither as electric charge nor as pure magnetic monopole \cite{key-32,key-33}
but with mixed behaviors of electric and magnetic charges rather namely
a tachyonic dyon. As such We agree with Negi-Rajput \cite{key-23}
that even in the case of a dyon interacting with generalized superluminal
electromagnetic fields, a tachyonic electric charge can not behave
as a bradyonic magnetic monopole or vice versa. Neither it behaves
exactly as dyons interacting with superluminal fields as the components
of electric and magnetic potential get mixed in different manner for
generalized superluminal electromagnetic fields. We do not have any
alternative left but to say that it is a kind tachyonic dyon interacting
with inconsistent natures of superluminal electromagnetic fields where
rotational ($curl$of vector potentials) counter parts of electric
and magnetic field do not occur. Transforming the force equation of
a dyon in frame $K$ i.e. 

\begin{eqnarray}
\overrightarrow{F} & = & e(\overrightarrow{E}+\overrightarrow{v}\times\overrightarrow{H})+g(\overrightarrow{H}-\overrightarrow{v}\times\overrightarrow{E}),\label{eq:36}\end{eqnarray}
we get the following equation of force in frame $K'$i.e.,

\begin{eqnarray}
\overrightarrow{F}' & = & e(\overrightarrow{E}'+\overrightarrow{v}'\times\overrightarrow{H}')+g(\overrightarrow{H}'-\overrightarrow{v}'\times\overrightarrow{E'}),\label{eq:37}\end{eqnarray}
under the mapping of fore said superluminal transformations in the
case of electric charge \cite{key-23} where the velocity becomes
the inverse velocity $\overrightarrow{v}'=\overleftarrow{\omega}=\frac{dt}{dz}$
. Equations (\ref{eq:35}) for superluminal electromagnetic fields
derived by using transformations (\ref{eq:24}) and corresponding
mappings are not consistent and do not describe the isotropic components
of electric and magnetic field vectors in all directions. On the other
hand, the components of a position four - vector become imaginary
in the direction perpendicular to relative motion between frame $K$
and $K'$. Similarly the perpendicular components of four - potential,
four - force, four - current and electromagnetic fields become imaginary
on passing from sub to superluminal realm via these transformations.
A lot of literature is also available \cite{key-17,key-34,key-35}
for the justification of imaginary superluminal transformations. Despite
of justifications, it is concluded that when we are prepared to consider
the tachyonic objects , we must give up the idea that dynamical quantities
or variables in classical mechanics are always real. 

To over come the various problems associated with both type of superluminal
Lorentz transformations, six - dimensional formalism \cite{key-36,key-37,key-38,key-39,key-40}
of space - time is adopted with the symmetric structure of space and
time having three space and three time components of a six dimensional
space time vector. The resulting space for bradyons and tachyons is
identified as the $R^{6}$- or $M(3,3)$ space where both space and
time and hence energy and momentum are considered as vector quantities.
Superluminal Lorentz transformations (SLTs) between two frames $K$
and $K'$ moving with velocity $v>1$ are defined in $R^{6}$- or
$M(3,3)$ space as follows;

\begin{eqnarray}
x\,' & = & \pm t_{x},\nonumber \\
y\,' & = & \pm t_{y},\nonumber \\
z\,' & = & \pm\gamma(z-vt),\nonumber \\
t_{x}\,' & = & \pm x,\nonumber \\
t_{y}\,' & = & \pm y,\nonumber \\
t_{z}\,' & = & \pm\gamma(t-vz).\label{eq:38}\end{eqnarray}
These transformations lead to the mixing of space and time coordinates
for transcendental tachyonic objects, $(|\overrightarrow{v}|\rightarrow\infty$
or $\overrightarrow{\omega}\rightarrow0)$ where equation (\ref{eq:38})
takes the following form;

\begin{eqnarray}
+\, dt_{x} & \rightarrow & dt_{x'}=dx\,+\nonumber \\
+\, dt_{y} & \rightarrow & dt_{y'}=dy\,+\nonumber \\
+\, dt_{z} & \rightarrow & dt_{z'}=dz\,+\nonumber \\
-\, dz & \rightarrow & dz\,'=dt_{x}\,-\nonumber \\
-\, dy & \rightarrow & dy\,'=dt_{y}\,-\nonumber \\
-\, dx & \rightarrow & dy\,'=dt_{z}\,-.\label{eq:39}\end{eqnarray}
It shows that we have only two four dimensional slices of $R^{6}$-
or $M\,(3,3)$ space $(+,+.+,-)$ and $(-,-,-,+)$. When any reference
frame describes bradyonic objects it is necessary to describe

\begin{eqnarray*}
M(1,3) & = & [t,x,y,z].\,\qquad(R^{4}-space)\end{eqnarray*}
So that the coordinates $t_{x}$ and $t_{y}$ are not observed or
couple together giving $t=(t_{x}^{2}+t_{y}^{2}+t_{z}^{2})^{\frac{1}{2}}$.
On the other hand when a frame describes bradyonic object in frame
$K$, it will describe a tachyonic object (with velocity $(|\overrightarrow{v}|\rightarrow\infty$
or $\overrightarrow{\omega}\rightarrow0)$ in $K'$ with $M'(1,3)$
space i.e.

\begin{eqnarray*}
M\,'(1,3) & = & [t_{z'},x',y',z']=[z,t_{x},t_{y},t_{z}].\,\qquad(T^{4}-space)\end{eqnarray*}
We define $M\,'(1,3)$ space as $T^{4}$- space or $M(3,1)$ space
where $x$ and $y$ are not observed or coupled together giving rise
to $r=(x^{2}+y^{2}+z)^{2\frac{1}{2}}$. As such, the spaces $R^{4}$
and $T^{4}$ are two observational slices of $R^{6}$- or $M(3,3)$
space but unfortunately the space is not consistent with special theory
of relativity. Subluminal and superluminal Lorentz transformations
loose their meaning in $R^{6}$- or $M(3,3)$ space with the sense
that these transformations do not represent either the bradyonic or
tachyonic objects in this space. It has been shown earlier \cite{key-23,key-24,key-25}
that the true localizations space for bradyons is $R^{4}$ - space
while that for tachyons is $T^{4}$ - space. So a bradyonic $R^{4}=M(1,3)$
space now maps to a tachyonic $T^{4}=M\,'(3,1)$ space or vice versa. 

\begin{eqnarray}
R^{4}=M(1,3) & \overset{SLT}{\rightarrow} & M\,'(3,1)=T^{4}.\label{eq:40}\end{eqnarray}
In a similar manner the corresponding mapping for the components of
electromagnetic potential in six - dimensional space may be written
as

\begin{eqnarray}
(\overrightarrow{V},i\phi) & \rightarrow & (\overrightarrow{\phi},iV)\label{eq:41}\end{eqnarray}
where

\begin{eqnarray*}
\phi=|\overrightarrow{\phi}| & = & (\phi_{x}^{2}+\phi_{y}^{2}+\phi_{z}^{2})^{\frac{1}{2}}\\
V=|\overrightarrow{V}| & = & (V_{x}^{2}+V_{y}^{2}+V_{z}^{2})^{\frac{1}{2}}.\end{eqnarray*}
The generalized four potential $\{\phi_{\mu}\}=\{\overrightarrow{\phi},iV\}$
associated with tachyonic dyon defined as

\begin{eqnarray}
\{\phi_{\mu}\} & = & \{\phi_{\mu}^{e}\}-i\{\phi_{\mu}^{m}\}\label{eq:42}\end{eqnarray}
where

\begin{eqnarray}
\{\phi_{\mu}^{e}\} & = & \{\overrightarrow{\phi^{e}},iA\},\nonumber \\
\{\phi_{\mu}^{m}\} & = & \{\overrightarrow{\phi^{m}},iB\}\label{eq:43}\end{eqnarray}
are the four - potentials associated with superluminal electric and
magnetic charges respectively with

\begin{eqnarray}
\phi^{e}=|\overrightarrow{\phi^{e}}| & = & (\phi_{x}^{e2}+\phi_{y}^{e2}+\phi_{z}^{e2})^{\frac{1}{2}},\nonumber \\
A & = & |\overrightarrow{A}|=(A_{x}^{2}+A_{y}^{2}+A_{z}^{2})^{\frac{1}{2}},\nonumber \\
\phi^{m}=|\overrightarrow{\phi^{m}}| & = & (\phi_{x}^{m2}+\phi_{y}^{m2}+\phi_{z}^{m2})^{\frac{1}{2}},\nonumber \\
B & = & |\overrightarrow{B}|=(B_{x}^{2}+B_{y}^{2}+B_{z}^{2})^{\frac{1}{2}}.\label{eq:44}\end{eqnarray}
Then the superluminal electric and magnetic fields of dyons in this
formalism will be described as 

\begin{eqnarray}
\overrightarrow{E}_{T} & = & -\overrightarrow{\nabla_{t}}A-\frac{\partial\overrightarrow{\phi_{e}}}{\partial r}-\overrightarrow{\nabla_{t}}\times\overrightarrow{\phi_{m}},\nonumber \\
\overrightarrow{H}_{T} & = & -\overrightarrow{\nabla_{t}}B-\frac{\partial\overrightarrow{\phi_{m}}}{\partial r}+\overrightarrow{\nabla_{t}}\times\overrightarrow{\phi_{e}}.\label{eq:45}\end{eqnarray}
The vector wave function $\overrightarrow{\psi_{T}}$ associated with
generalized electromagnetic fields in superluminal transformation
is defined as 

\begin{eqnarray}
\overrightarrow{\psi_{T}} & = & \overrightarrow{E_{T}}-i\overrightarrow{H_{T}}.\label{eq:46}\end{eqnarray}
Then we get the following pair of generalized Maxwell's equation for
generalized fields of dyons in $T^{4}$- space (for tachyonic dyons
via superluminal transformation) i.e.

\begin{eqnarray}
\overrightarrow{\nabla_{t}}.\overrightarrow{\psi_{T}} & = & \Im_{0}\nonumber \\
\overrightarrow{\nabla_{t}}\times\overrightarrow{\psi_{T}} & = & -i\overrightarrow{\Im}-i\frac{\partial\overrightarrow{\psi_{T}}}{\partial r}\label{eq:47}\end{eqnarray}
where $\Im_{0}$ and $\overrightarrow{\Im}$ are the components of
generalized four - current source densities of dyons in $T^{4}$ -
space , given by 

\begin{eqnarray}
\{\rho_{\mu}\} & = & \{\rho_{\mu}^{e}\}-i\{\rho_{\mu}^{m}\}=\{\Im_{0},\overrightarrow{\Im}\}.\label{eq:48}\end{eqnarray}
Substituting relation (\ref{eq:45}) into equation (\ref{eq:46})
as we have done earlier and using equation (\ref{eq:44}), we obtain
the following expression for generalized vector field in terms of
the components of generalized four potential of dyon in $T^{4}$ -
space i.e.

\begin{eqnarray}
\overrightarrow{\psi_{T}} & = & -\frac{\partial\overrightarrow{\phi}}{\partial r}-\overrightarrow{\nabla_{t}}V-i\overrightarrow{\nabla_{t}}\times\overrightarrow{\phi}.\label{eq:49}\end{eqnarray}
As such, we can write the following tensorial forms of generalized
Maxwell's - Dirac equations of dyons under the influence of superluminal
transformation (in $T^{4}$- space) i.e.

\begin{eqnarray}
f_{\mu\nu,\nu} & = & \rho_{\mu}^{e},\nonumber \\
\widetilde{f_{\mu\nu,\nu}} & = & \rho_{\mu}^{m}\label{eq:50}\end{eqnarray}
where $f_{\mu\nu}$ and $\widetilde{f_{\mu\nu}}$ are described (like
equations (\ref{eq:10})) as superluminal electric and magnetic fields
tensors in $T^{4}$ - space giving rise to Maxwell's equations coupled
to electric and magnetic four currents in $T^{4}$- space.With the
help of equation (\ref{eq:47}) we may obtain a new vector parameter
(current) $\overrightarrow{S_{T}}$, in $T^{4}$- space i.e.

\begin{eqnarray}
\overrightarrow{S_{T}} & = & \square_{t}\overrightarrow{\psi_{T}}=\frac{\partial\overrightarrow{\Im}}{\partial r}-\overrightarrow{\nabla_{t}}\Im_{0}-i\overrightarrow{\nabla_{t}}\times\overrightarrow{\Im}\label{eq:51}\end{eqnarray}

where

\begin{eqnarray}
\square_{t} & = & \partial_{r}^{2}-\nabla_{t}^{2}=\frac{\partial}{\partial r^{2}}-\frac{\partial}{\partial t_{x}^{2}}-\frac{\partial}{\partial t_{y}^{2}}-\frac{\partial}{\partial t_{z}^{2}}.\label{eq:52}\end{eqnarray}
We may also define the generalized field tensor for tachyonic dyons
associated with generalized superluminal electromagnetic fields in
$T^{4}$- space as

\begin{eqnarray}
g_{\mu\nu} & = & f_{\mu\nu}-i\,\widetilde{f_{\mu\nu}}.\label{eq:53}\end{eqnarray}
It gives directly the following form of field equation (parallel to
Maxwell's equations associated with generalized superluminal electromagnetic
fields of dyons) in $T^{4}$- space i.e.

\begin{eqnarray}
g_{\mu\nu,\nu} & = & \rho_{\mu},\nonumber \\
\widetilde{g_{\mu\nu,\nu}} & = & 0\label{eq:54}\end{eqnarray}
where $g_{\mu\nu}=\phi_{\mu,\nu}-\phi_{\nu,\mu}$ is denoted as the
generalized electromagnetic field tensor and $\rho_{\mu}=\rho_{\mu}^{e}-i\rho_{\mu}^{m}$
is described as the generalized four current associated with superluminal
electromagnetic fields of dyons in $T^{4}$- space. Equation (\ref{eq:54})
may also be written as follows ( like second order Klein-Gordon equation),
for dyonic fields in $T^{4}$- space, i.e.

\begin{eqnarray}
\square_{t}\phi_{\mu} & = & \rho_{\mu}\qquad(in\, Lorentz\, gauge).\label{eq:55}\end{eqnarray}
Lorentz gauge condition and continuity equation in $T^{4}$- space
are written as

\begin{eqnarray}
\frac{\partial V}{\partial r}+\overrightarrow{\nabla_{t}}.\overrightarrow{\phi} & = & 0\,\qquad(Lorentz\, gauge\, condition)\label{eq:56}\\
\frac{\partial J}{\partial r}+\overrightarrow{\nabla_{t}}.\overrightarrow{\rho} & = & 0\,\qquad(Continuity\, equation).\label{eq:57}\end{eqnarray}
The suitable manifestly covariant Lagrangian density, which yields
the field equation (\ref{eq:40}) under the variation of field parameters
i.e. potential only without changing the trajectory of particle, may
be written as follows in $T^{4}$- space;

\begin{eqnarray}
L_{T} & = & -m_{0}-\frac{1}{4}g_{\mu\nu}g_{\mu\nu}^{\star}+\phi_{\mu}^{\star}\rho_{\mu}\label{eq:58}\end{eqnarray}
where $m_{0}$ is the rest mass of the particle, $(\star)$ denotes
the complex conjugate and

\begin{eqnarray}
\widetilde{g_{\mu\nu}} & = & \frac{1}{2}\varepsilon_{\mu\nu\rho\sigma}g^{\rho\sigma}.\label{eq:59}\end{eqnarray}
Lagrangian density given by equation (\ref{eq:58}) directly yields
the following Lorentz force equation of dyons under superluminal transformations
i.e.

\begin{eqnarray}
\zeta_{\mu} & = & m_{0}\ddot{x_{\mu}=Re(q*g_{\mu\nu})u^{\nu}}\label{eq:60}\end{eqnarray}
where $q$ is the generalized charge of dyon, $Re$ denotes real part,
$\ddot{x_{\mu}}$ is the four acceleration and $u^{\nu}$ is the inverse
velocity of the particle in $T^{4}$- space and is given by,

\begin{eqnarray*}
u^{\nu} & = & \frac{dt^{\nu}}{dz}\end{eqnarray*}
or

\begin{eqnarray}
\overleftarrow{u} & = & (\frac{dt_{x}}{dz},\frac{dt_{y}}{dz},\frac{dt_{z}}{dz}).\label{eq:61}\end{eqnarray}
Equation (\ref{eq:60}) describes the following form of Lorentz force
for dyons interacting with superluminal electromagnetic fields in
$T^{4}$- space,

\begin{eqnarray}
\overrightarrow{\zeta_{\mu}} & = & e(\overrightarrow{E_{T}}+\overleftarrow{u}\times\overrightarrow{H_{T}})+g(\overrightarrow{H_{T}}-\overleftarrow{u}\times\overrightarrow{E_{T}}).\label{eq:62}\end{eqnarray}

\section{Quaternionic Formulation of Dyons in $T^{4}$- space}

A quaternion $q$ is defined as

\begin{eqnarray}
q & = & q_{0}e_{0}+q_{1}e_{1}+q_{2}e_{2}+q_{3}e_{3}\label{eq:63}\end{eqnarray}
where $q_{\alpha}(\alpha=0,1,2,3)$ are real or complex numbers with
three imaginary units $e_{1},\, e_{2}$and $e_{3}$ satisfy the relations,

\begin{eqnarray}
e_{i}e_{j} & = & -\delta_{ij}+\varepsilon_{ijk}e_{k}.\label{eq:64}\end{eqnarray}
All laws of algebra with the exception of commutative law of multiplication
are satisfied by quaternions, which form a division ring. The set
of quaternions is called a $4$ - dimensional real space essentially
because $4$- real numbers are required to specify a quaternion. The
complex conjugate of $q$ i.e. $q^{\star}$ is defined by

\begin{eqnarray}
q^{\star} & = & q_{0}+q_{1}^{\star}e_{1}+q_{2}^{\star}e_{2}+q_{3}^{\star}e_{3}\label{eq:65}\end{eqnarray}
and the quaternion conjugate of $q$ by

\begin{eqnarray}
\overline{q} & = & q_{0}-q_{1}e_{1}-q_{2}e_{2}-q_{3}e_{3}\label{eq:66}\end{eqnarray}
which gives

\begin{eqnarray}
(qp)^{\star} & = & q^{\star}p^{\star}\nonumber \\
(\overline{qp}) & = & \overline{p}\,\overline{q}\label{eq:67}\end{eqnarray}
where $p$ and $q$ , are also the quaternions showing that the quaternion
conjugate of product of two quaternions is the product of conjugates
in reverse order.

The scalar product of a two quaternions is defined as

\begin{eqnarray*}
(p,q) & = & \frac{1}{2}(p\overline{q}+q\overline{p})\end{eqnarray*}
and the norm of a quaternion is given as

\begin{eqnarray}
|p| & =(p.p)= & p_{0}^{2}+p_{1}^{2}+p_{2}^{2}+p_{3}^{2}.\label{eq:68}\end{eqnarray}
The inverse of a quaternion $q$is also a quaternion and given by

\begin{eqnarray}
q^{-1} & = & \frac{q}{|q|^{2}}.\label{eq:69}\end{eqnarray}
As such, it is easy to write a four vector in $T^{4}$- space as a
quaternion. Adopting the same procedure to write the quantum equation
in quaternion formalism, (as we have done earlier \cite{key-41,key-42}),
the generalized four-potential and generalized four - current associated
with dyons under superluminal Lorentz transformation ($T^{4}$ -space)
may be written as quaternions i.e.

\begin{eqnarray}
\phi & = & -iV+e_{1}\phi_{1}+e_{2}\phi_{2}+e_{3}\phi_{3}\label{eq:70}\\
\rho & = & -iJ+e_{1}\rho_{1}+e_{2}\rho_{2}+e_{3}\rho_{3}.\label{eq:71}\end{eqnarray}
In this case quaternionic differential operator is written as 

\begin{eqnarray}
\boxdot_{t} & = & -i\partial_{r}+e_{1}\partial_{1}+e_{2}\partial_{2}+e_{3}\partial_{3}\label{eq:72}\end{eqnarray}
where

\begin{eqnarray}
\partial_{r}=\frac{\partial}{\partial r}\,, & \partial_{1}=\frac{\partial}{\partial t_{x}}, & \partial_{2}=\frac{\partial}{\partial t_{y}},\partial_{3}=\frac{\partial}{\partial t_{z}}.\label{eq:73}\end{eqnarray}
Operating equation (\ref{eq:72}) on equations (\ref{eq:70}) and
(\ref{eq:71}) and using equations (\ref{eq:49}) ,(\ref{eq:56})
and (\ref{eq:57}), we get;

\begin{eqnarray}
\boxdot_{t}\phi & = & -(\partial_{r}V+\partial_{1}\phi_{1}+\partial_{2}\phi_{2}+\partial_{3}\phi_{3})\nonumber \\
-i & e_{1} & \{-\partial_{r}\phi_{1}-\partial_{1}V-i(\partial_{2}\phi_{3}-\partial_{3}\phi_{2})\}\nonumber \\
-i & e_{2} & \{-\partial_{r}\phi_{2}-\partial_{2}V-i(\partial_{3}\phi_{1}-\partial_{1}\phi_{3})\}\nonumber \\
-i & e_{3} & \{-\partial_{r}\phi_{3}-\partial_{3}V-i(\partial_{1}\phi_{2}-\partial_{2}\phi_{1})\}\nonumber \\
= & -\psi_{r} & -i(e_{1}\psi_{1}+e_{2}\psi_{2}+e_{3}\psi_{3})\label{eq:74}\end{eqnarray}
and

\begin{eqnarray}
\boxdot_{t}\rho & = & -(\partial_{r}J+\partial_{1}\rho_{1}+\partial_{2}\rho_{2}+\partial_{3}\rho_{3})\nonumber \\
-i & e_{1} & \{-\partial_{r}\rho_{1}-\partial_{1}J-i(\partial_{2}\rho_{3}-\partial_{3}\rho_{2})\}\nonumber \\
-i & e_{2} & \{-\partial_{r}\rho_{2}-\partial_{2}J-i(\partial_{3}\rho_{1}-\partial_{1}\rho_{3})\}\nonumber \\
-i & e_{3} & \{-\partial_{r}\rho_{3}-\partial_{3}J-i(\partial_{1}\rho_{2}-\partial_{2}\rho_{1})\}\nonumber \\
= & -S_{r} & -i(e_{1}S_{1}+e_{2}S_{2}+e_{3}S_{3})\label{eq:75}\end{eqnarray}
where

\begin{eqnarray}
\psi_{r} & = & \partial_{r}V+\partial_{1}\phi_{1}+\partial_{2}\phi_{2}+\partial_{3}\phi_{3}=0\:(Lorentz\,\, Gauge\,\, condition)\label{eq:76}\end{eqnarray}

\begin{eqnarray}
S_{r} & = & \partial_{r}J+\partial_{1}\rho_{1}+\partial_{2}\rho_{2}+\partial_{3}\rho_{3}=0.\quad(Continuity\, equation)\label{eq:77}\end{eqnarray}
We may then write equations (\ref{eq:74}) and (\ref{eq:75}) in the
following quaternionic forms;

\begin{eqnarray}
\boxdot_{t}\phi & = & \psi_{T}\label{eq:78}\end{eqnarray}
and

\begin{eqnarray}
\boxdot_{t}\rho & = & S_{T}.\label{eq:79}\end{eqnarray}
These are the quaternionic differential equations for generalized
potential and generalized current in $T^{4}$- space under superluminal
Lorentz transformations. The conjugate representation of quaternion
field equations (\ref{eq:78}) and (\ref{eq:79}) in $T^{4}$ - space
under superluminal Lorentz transformations may then be expressed as,

\begin{eqnarray}
\overline{\boxdot_{t}}\,\overline{\phi} & = & \overline{\psi_{T}}\label{eq:80}\\
\overline{\boxdot_{t}}\,\overline{\rho} & = & \overline{S_{T}}\label{eq:81}\end{eqnarray}
where $\overline{\boxdot_{t}},\,\overline{\phi},\,\overline{\rho},\:\overline{\psi_{T}}$
and $\overline{S_{T}}$ are the quaternion conjugate and defined as,

\begin{eqnarray}
\overline{\boxdot_{t}} & = & -i\partial_{r}-(e_{1}\partial_{1}+e_{2}\partial_{2}+e_{3}\partial_{3})\label{eq:82}\\
\overline{\phi} & = & -iV-(e_{1}\phi_{1}+e_{2}\phi_{2}+e_{3}\phi_{3})\label{eq:83}\\
\overline{\rho} & = & -iJ-(e_{1}\rho_{1}+e_{2}\rho_{2}+e_{3}\rho_{3})\label{eq:84}\\
\overline{\psi_{T}} & = & -\psi_{r}+i(e_{1}\psi_{1}+e_{2}\psi_{2}+e_{3}\psi_{3})\label{eq:85}\\
\overline{S_{T}} & = & -S_{r}-i(e_{1}S_{1}+e_{2}S_{2}+e_{3}S_{3}).\label{eq:86}\end{eqnarray}
Similarly, we may derive the following quaternionic forms of other
fields of dyons in $T^{4}$ - space under superluminal Lorentz transformation
given by equations (\ref{eq:51}), (\ref{eq:55}) (\ref{eq:56}) and
(\ref{eq:57}) as

\begin{eqnarray}
\overline{\boxdot_{t}}\boxdot_{t}\psi_{T} & = & -S_{T}\label{eq:87}\\
\overline{\boxdot_{t}}\boxdot_{t}\phi & = & -\rho\label{eq:88}\\
{}[\boxdot_{t},g] & = & \rho\label{eq:89}\\
q[v,g] & = & f\label{eq:90}\end{eqnarray}
where

\begin{eqnarray}
v & = & -iv_{0}e_{0}+e_{1}v_{1}+e_{2}v_{2}+e_{3}v_{3},\nonumber \\
g & = & -ig_{0}e_{0}+e_{1}g_{1}+e_{2}g_{2}+e_{3}g_{3},\nonumber \\
f & = & -if_{0}e_{0}+e_{1}f_{1}+e_{2}f_{2}+e_{3}f_{3}.\label{eq:91}\end{eqnarray}
In equation (\ref{eq:90}) $v,\: g$ and $f$ are respectively the
quaternionic forms of inverse velocity, generalized field tensor and
Lorentz force associated with dyons in $T^{4}$- space under superluminal
Lorentz transformations.

As such, the tensorial forms of generalized field equations (\ref{eq:89})
and (\ref{eq:90}) are analogous to the following quaternionic forms;

\begin{eqnarray}
[\boxdot_{t},g_{\mu}] & = & \rho_{\mu}\,(\mu=0,1,2,3)\label{eq:92}\end{eqnarray}
and

\begin{eqnarray}
q[v,g_{\mu}] & = & f_{\mu}\label{eq:93}\end{eqnarray}
where $\rho_{\mu}$ and $f_{\mu}$ are the four - current and four
- force associated with generalized fields of dyons in $T^{4}$- space
superluminal Lorentz transformation. The norms of quaternions $\boxdot_{t},\,\phi$
and $\rho$ may then be obtained as 

\begin{eqnarray}
\boxdot_{t}^{2} & = & \frac{1}{2}[\boxdot_{t},\boxdot_{t}]=\overline{\boxdot_{t}}\boxdot_{t}=-\partial_{r}^{2}+|\nabla_{t}|^{2}=-\square_{t}=-\left|\boxdot\right|^{2};\label{eq:94}\\
\phi^{2} & = & =\frac{1}{2}[\phi,\phi]=\overline{\phi}\phi=-V^{2}+|\phi|^{2}=-|\phi|^{2};\label{eq:95}\\
\rho^{2} & = & =\frac{1}{2}[\rho,\rho]=\overline{\rho}\rho=-J^{2}+|\rho|^{2}=-|\rho|^{2}\label{eq:96}\end{eqnarray}
where $\overline{\boxdot_{t}},\,\overline{\phi}$and $\overline{\rho}$
are Hamiltonian conjugate of $\boxdot_{t},\,\phi$ and $\rho$ respectively.
We may thus see that under the influence of quaternions the norm of
four vectors is changed like the norms of four vectors do under imaginary
superluminal transformations. The same conclusion may be drawn for
the generalized form of Maxwell's equations for dyon on passing from
subluminal and superluminal electromagnetic fields. It may therefore
be concluded that the quaternionic forms of field equation may be
regarded as the Maxwell's equations under the influence of imaginary
superluminal transformations, which lead to the mapping $(3,1)\longleftrightarrow(1,3)$
of space-time.

As such, from the fore going analysis we may again draw the same conclusion,
also for the theory of generalized fields of dyons , as it has already
be drawn \cite{key-41} that the complex quantum mechanics for time-like
particles (bradyons) in subluminal frame of reference reduces to quaternion
quantum mechanics for space-like particles (tachyons) in superluminal
frame of reference or vice versa. The advantage in expressing the
field equations in quaternionic form is that one may extend the theory
of bradyons (the Cauchy data at $t=0$) to the theory of tachyons
(the Cauchy data at $r=0$) directly in this formalism \cite{key-43}
and accordingly the space-time duality and space-time reciprocity
be tackled between complex and quaternion quantum mechanics \cite{key-44}.
Finally, it may be pointed out that relativistic equations in quaternionic
form will describe the theory of both bradyons tachyons only when
making the use of complex (bi) quaternions. The quaternionic formalism
described here is thus compact, simpler, unique and consistent. It
is also manifestly covariant under quaternion Lorentz transformations.

\begin{description}
\item [{Acknowledgment-}]~
\end{description}
One of us OPSN is thankful to Chinese Academy of Sciences and Third
world Academy of Sciences for awarding him CAS-TWAS visiting scholar
fellowship to pursue a research program in China. He is also grateful
to Professor Tianjun Li for his hospitality at Institute of Theoretical
Physics, Beijing, China.

\end{document}